\documentclass[aps,longbibliography,
  twocolumn,amsmath,
  amssymb,nofootinbib, superscriptaddress, preprintnumbers]{revtex4-1}

\usepackage{graphics}      % standard graphics specifications
\usepackage{graphicx}      % alternative graphics specifications
\usepackage{longtable}     % helps with long table options
\usepackage{url}           % for on-line citations
\usepackage{bm}            % special 'bold-math' package
\usepackage{xcolor}
\usepackage{mathrsfs}
\usepackage[colorlinks,allcolors=blue]{hyperref}
\usepackage{blindtext}
\usepackage{hyperref}
\usepackage{float}
\usepackage{bbold}
\usepackage{lipsum}
\usepackage[normalem]{ulem}
\usepackage{orcidlink}
\usepackage{academicons}
\usepackage{xcolor}
\usepackage{subfigure}
\usepackage{booktabs}
\usepackage{hyperref}

\newcommand{\orcid}[1]{\href{https://orcid.org/#1}{\textcolor[HTML]{A6CE39}{\aiOrcid}}}

\begin{document}
\preprint{MPP-2023-237}

\title{Detecting Fast Neutrino Flavor Conversions with Machine Learning}
%: the Possibility of the Occurrence of Fast Neutrino Flavor Conversion Modes}

\newcommand*{\MPP}{\textit{\small{Max-Planck-Institut f\"ur Physik (Werner-Heisenberg-Institut), F\"ohringer Ring 6, 80805 M\"unchen, Germany}}}

\author{Sajad Abbar \orcidlink{0000-0001-8276-997X}   } 
\affiliation{\MPP}
\author{Hiroki Nagakura \orcidlink{0000-0002-7205-6367}} 
\affiliation{Division of Science, National Astronomical Observatory of Japan, 2-21-1 Osawa, Mitaka, Tokyo 181-8588, Japan}

%\date{\today}

\begin{abstract}
Neutrinos in dense environments like core-collapse supernovae (CCSNe) and neutron star mergers (NSMs) can 
undergo fast flavor conversions (FFCs) once the angular distribution of neutrino  lepton number   crosses zero along a certain direction. Recent advancements have demonstrated the effectiveness of machine learning (ML) in detecting these  crossings.
In this study, we enhance prior research in two significant ways. Firstly, we utilize realistic data from CCSN simulations, where neutrino transport is solved using the full Boltzmann equation. We evaluate the ML methods' adaptability in a real-world context, enhancing their robustness. 
In particular, we demonstrate that when working with artificial data, simpler models outperform their more complex counterparts, a noteworthy illustration of the bias-variance tradeoff in the context of ML.
We also explore methods to improve artificial datasets for ML training.
In addition, we extend our ML techniques to detect the crossings in the heavy-leptonic channels, accommodating scenarios where $\nu_x$ and $\bar\nu_x$ may differ. 
Our research highlights the extensive versatility and effectiveness of ML techniques, presenting an unparalleled opportunity to evaluate the occurrence of FFCs in CCSN and NSM simulations.

 \end{abstract}

\maketitle

%%%%%%%%%%%%%%%%%%%%%%%%%%%%%%%%%%%%%%%%%%%%%%%%%%%%%%%%%%%%%%%%%%%%%%%%%%%%%
\section{Introduction}

Core-collapse supernovae (CCSNe) and neutron star mergers (NSMs) are cataclysmic stellar events that represent the dramatic culmination of massive stars' life cycles and the collision and coalescence of incredibly dense remnants, respectively~\cite{Burrows:2020qrp, Janka:2012wk, Foucart:2022bth, Kyutoku:2021icp}. These events not only mark the end of massive stars and dense objects, but also unveil some of the most energetic and enigmatic phenomena in the universe.
In the heart of these cosmic fireworks, one of the most fascinating processes at play is the neutrino emission, which are released in vast quantities during CCSNe and NSMs.

As they journey through the extraordinarily dense and extreme conditions within these events, neutrinos undergo an intriguing phenomenon known as collective neutrino oscillations.
This fascinating behavior arises from their interactions with the dense background neutrino gas, where coherent forward scatterings play a pivotal role. This phenomenon 
occurs in a nonlinear and collective manner, creating a rich tapestry of flavor transformations~\cite{pantaleone:1992eq, sigl1993general, Pastor:2002we,duan:2006an, duan:2006jv, duan:2010bg, Mirizzi:2015eza}~(for a recent review see Ref.~\cite{Volpe:2023met}).

Of particular interest are the so-called \emph{fast} flavor conversions (FFCs), which occur on scales characterized by $\sim G_{\rm{F}}^{-1} n_{\nu}^{-1}$ (see, e.g., Refs.~\cite{Sawyer:2005jk, Sawyer:2015dsa,
Chakraborty:2016lct, Izaguirre:2016gsx,Capozzi:2017gqd, Abbar:2017pkh, Abbar:2018beu,Capozzi:2018clo, Martin:2019gxb, Abbar:2018shq, Abbar:2019zoq, Capozzi:2019lso, Johns:2019izj, Martin:2021xyl, Tamborra:2020cul,  Sigl:2021tmj, Kato:2021cjf,  Morinaga:2021vmc, Nagakura:2021hyb,  Sasaki:2021zld, Padilla-Gay:2021haz, Abbar:2020qpi, Capozzi:2020syn, DelfanAzari:2019epo, Harada:2021ata,  Abbar:2021lmm, Just:2022flt, 
Padilla-Gay:2022wck, Capozzi:2022dtr, Zaizen:2022cik, Shalgar:2022rjj,  Kato:2022vsu, Zaizen:2022cik,  Bhattacharyya:2020jpj, Wu:2021uvt, Richers:2021nbx, Richers:2021xtf, Dasgupta:2021gfs, Nagakura:2022kic, Ehring:2023lcd, Ehring:2023abs, Xiong:2023vcm, Zaizen:2023ihz, Xiong:2023upa, Fiorillo:2023hlk, Nagakura:2023wbf, Martin:2023gbo, Fiorillo:2023mze,
Grohs:2023pgq}). Here, $G_{\rm{F}}$ represents the Fermi coupling constant, and $n_{\nu}$ denotes the neutrino number density. These FFCs can take place on timescales much shorter than what would be expected in the vacuum.

FFCs occur \emph{iff} the angular distribution of the neutrino lepton
number,  defined as,
\begin{equation}
\begin{split}
  G(\mathbf{v}) =
  \sqrt2 G_{\mathrm{F}}
  \int_0^\infty  \frac{E_\nu^2 \mathrm{d} E_\nu}{(2\pi)^3}
        &[\big( f_{\nu_e}(\mathbf{p}) -  f_{\nu_x}(\mathbf{p}) \big)\\
              &- \big( f_{\bar\nu_e}(\mathbf{p}) -  f_{\bar\nu_x}(\mathbf{p}) \big)],
 \label{Eq:G}
\end{split}
\end{equation}
crosses zero at some $\mathbf{v} = \mathbf{v}(\mu,\phi_\nu)$, with $\mu =\cos\theta_\nu$~\cite{Morinaga:2021vmc}. 
Here, $E_\nu$, $\theta_\nu$, and $\phi_\nu$ are the neutrino energy,  
the zenith, and azimuthal angles of the neutrino velocity, respectively, 
and  $f_{\nu}$'s are the neutrino 
occupation numbers.
When  $\nu_x$ and $\bar\nu_x$ have similar angular distributions, a scenario commonly observed in state-of-the-art 
CCSN 
% {\bf CCSN}
simulations, this expression transforms into the conventional $\nu$ELN (neutrino electron lepton number).

Exploring $\nu$ELN crossings necessitates access to the complete angular distributions of neutrinos. However, obtaining such detailed angular information poses a significant challenge in most cutting-edge CCSN and NSM simulations due to the prohibitive computational demands involved.

As a practical alternative, many simulations simplify neutrino transport by relying on a limited set of angular distribution
 moments~\cite{Shibata:2011kx, Cardall:2012at, thorne1981relativistic}. In our specific investigation, we focus on radial moments, defined as,
\begin{equation}
I_n = \int_{-1}^{1} \mathrm{d}\mu\ \mu^n\ \int_0^\infty \int_0^{2\pi} \frac{E_\nu^2 \mathrm{d} E_\nu \mathrm{d} \phi_\nu}{(2\pi)^3} \
        f_{\nu}(\mathbf{p}).
\end{equation}
These moments capture the key aspects of the neutrino angular distribution, at the same time allowing for its more computationally manageable treatment.

Note that our current emphasis lies on axisymmetric crossings, directing our attention specifically to radial moments where the angular distribution integrates over $\phi_\nu$. It is important to note that non-axisymmetric crossings fall outside the scope of our present study, a matter to be explored in future works.

 %For instance,  the $M_1$ closure scheme directly tracks the evolution of two moments, namely $I_0$ and $I_1$. 
%This choice significantly reduces computational overhead. Importantly, the moments $I_2$ and $I_3$ are related to $I_0$ and $I_1$ through analytical closure relations, further streamlining the simulation process. This strategic simplification enables us to gain valuable insights into the physics of $\nu$ELN crossings while maintaining computational feasibility.

Despite the inherent loss of a significant amount of information when considering only a select few neutrino angular moments, ingenious methods can still be devised to harness these limited information for assessing FFCs in CCSN and NSM simulations.
In the initial stages of research in this field, the primary focus was on analytical or semi-analytical
 techniques~\cite{Dasgupta:2018ulw, Abbar:2020fcl, Johns:2021taz, Johns:2019izj, Nagakura:2021hyb, Richers:2022dqa, Nagakura:2021suv}. While these methods have demonstrated their ability to capture ELN crossings and have found relative success in the literature, they are constrained in their performance.
This limitation arises from their either sluggish computational speed or their inefficiency in identifying ELN crossings which impacts their ability to efficiently detect FFCs in real-time simulations. Specifically, the most efficient techniques tend to be noticeably sluggish, and their development can be relatively intricate when starting from scratch~\footnote{As discussed in the text, the traditional methodologies can be broadly categorized into two classes: analytical methods, which concentrate on the instability of a specific mode, and semi-analytical models, which focus on identifying descriptive polynomials or fitting functions to angular distributions. While analytical approaches boast speed and ease of implementation, their efficiency in the CCSN environment is generally questioned due to specific conditions required for their usefullness. Conversely, semi-analytical methods, while being more efficient in capturing crossings, pose a significant computational burden. In contrast,  ML, by focusing on simulation-derived moments (it can only focus on $I_0$ and $I_1$), presents a promising alternative that mitigates the influence of variations and artefacts introduced by closure relations. This should be compared with the results, e.g. in Ref.~\cite{Abbar:2020fcl}, that showed
using the semi-analytical methods could only capture $\lesssim 50\%$ of the ELN crossings (although more angular information were provided, namely $I_2$ and $I_3$). Therein, it was also shown that the simple method developed in Ref.~\cite{Dasgupta:2018ulw} could not capture any ELN crossings.}.

Recent research has demonstrated the  effectiveness of machine learning (ML) techniques in identifying FFC in CCSN and NSM simulations~\cite{Abbar:2023kta}. While ML methods are data-intensive and require an initial training phase with data, it's important to note that once trained, they exhibit exceptional speed and efficiency. This presents a promising avenue for real-time detection of FFI's within the context of CCSN and NSM simulations.
Moreover, integrating pre-trained ML models is a straightforward procedure, significantly reducing the requirement for extensive coding work, even when analyzing the occurrence of FFCs in a post-processing phase. In fact, ML techniques offer the fastest and most precise approach to detect FFCs, and their performance in more complex environments can be further enhanced as and if one can train them on the data obtained from those environments.

 In this paper, we advance the prior study in two pivotal directions.
Firstly, in earlier work, ML models were trained using artificial data generated from specific parametric angular distributions. While these models showed promise, they were only partially validated against a limited amount of realistic data from NSM remnant simulations. It is essential for ML techniques to be trained and tested on data that closely resembles real-world 
simulations~\footnote{Henceforth in this paper, whenever the term \emph{realistic} or \emph{real-world data} is mentioned, it specifically refers to the data acquired from our axisymmetric CCSN simulation.}. In our study, we take a significant step forward by utilizing authentic data from a CCSN simulation, where neutrino transport was  modeled using the full Boltzmann equation. This  allows us to assess the adaptability of ML methods in a real-world context and examine their limitations as well as their optimal performance range.
Furthermore, we acknowledge that artificial data are more readily available and can offer broader distributions that are expected in CCSN and NSM environments. Consequently, we also investigate methods to enhance artificial datasets for ML training purposes.

In addition, in our previous study, we assumed that the distributions of $\nu_x$ and $\bar\nu_x$ were identical. From a  ML perspective, this simplification facilitated the development of our ML module by requiring a smaller number of features and a more efficient  classification. 
In this work, we develop ML techniques to detect the crossings regarding the neutrino heavy-leptnic channel distribution ($\nu$XLN) addressing scenarios where $\nu_x$ and $\bar\nu_x$ may exhibit differences, a \emph{previously unexplored} area in the literature. While our results may be slightly less accurate than those in the previous scenario, ML methods still prove remarkably effective in identifying the occurrence of FFC in this scenario..

In the upcoming section, we delve into our CCSN model, the source of our data. We then assess the performance of ML methods in detecting FFCs in our CCSN model, specifically focusing on the detection of $\nu$ELN crossings. Finally and before presenting our conclusions, we analyze the performance of ML methods in detecting the crossings in the heavy-leptnic channel.

  \section{CCSN model}\label{sec:SN}
 Here, we construct a ML technique to detect $\nu$ELN- and XLN angular crossings based on an axisymmetric CCSN model with full Boltzmann neutrino transport~\cite{Nagakura:2019evv}. Before entering into the detail of our ML, we briefly describe our CCSN model providing neutrino dataset used for ML training and its testing.

The numerical simulation was carried out by a Boltzmann-neutrino-radiation hydrodynamic code~\cite{Nagakura:2014nta} with some special treatments to handle proper motions of proto-neutron star (PNS)~\cite{Nagakura:2016jcl,Nagakura:2019rdf}. In this model, a table of multi-nuclear variational method equation-of-state~\cite{Furusawa:2017auz}  was used, and the nuclear abandance in the table was also used to compute neutrino-matter interactions for the consistent treatment between EOS and weak rates~\cite{Nagakura:2018qpg}. We used a 11.2 solar mass progenitor model in Ref.~\cite{Woosley:2002zz}.

One of the noticeable features in the CCSN model is that large-scale asymmetric neutrino emission emerges $>150$ms after bounce, that corresponds to the timing of asymmetric shock expansion and the onset of PNS proper motion (see Fig. 1 in Ref.~\cite{Nagakura:2019evv}). We note that the asymmetric emission is clearly anti-correlated between $\nu_e$ and $\bar{\nu}_e$, which is attributed to the distribution of electron fraction ($Y_e$) in the vicinity of PNS, and $\bar{\nu}_e$ tends to be more abundant in low $Y_e$ environments. As shown in Ref.~\cite{Nagakura:2019sig}, the increase of $\bar{\nu}_e$ reduces the disparity between $\nu_e$ and $\bar{\nu}_e$ angular distributions, leading to enhance the possibility of $\nu$ELN crossings. In the region with higher $Y_e$ environments, on the other hand, $\nu_e$ becomes much higher than $\bar{\nu}_e$, indicating that FFI is unlikely to occur. 

In the CCSN model, $\nu$ELN angular crossings are observed rather stably at $>200$ms (see Fig. 2 in Ref.~\cite{Nagakura:2019sig}); hence, we employ three different time snapshots for our ML training (200, 250, and 300ms after bounce), in which there are both spatial regions with and without $\nu$ELN angular crossings. It should also be mentioned that $\nu$ELN crossings are also observed in PNS convective layer  and in pre-shock regions, which are consistent with previous studies~\cite{Glas:2019ijo, DelfanAzari:2019tez, Morinaga:2019wsv}. For more detailed discussion about neutrino angular distributions associated with arguments of $\nu$ELN crossings, we refer readers to Ref.~\cite{Nagakura:2019sig}.

 %We discuss the SN model..... 
 %\textcolor{red}{I suggest we show the plots of axisymmetric ELN crossings and discuss them.
 %I also suggest we do not discuss non-axisymmetric symmetric crossings here and postpone it to the conclusion
 %to avoid the referee questioning and also to motivate our next paper}

   \section{ML algorithms}\label{sec:ML}
   
   ML, at the crossroads of computer science and artificial intelligence, is transforming how computers learn and make decisions from data. 
   By unraveling intricate patterns, ML drives progress across diverse domains, from image and speech recognition to healthcare, finance, and autonomous vehicles. 
   
Recent advancements have demonstrated the effective utilization of ML algorithms for detecting $\nu$ELN crossings in CCSN and NSM simulations~\cite{Abbar:2023kta}. In this context, we commence with a brief overview of the data preparation and ML techniques employed in Ref.~\cite{Abbar:2023kta}. Subsequently, we provide a comprehensive discussion of our research findings.

To effectively train and evaluate our ML algorithms, it is imperative to possess a substantial dataset comprising labeled values for $I_0$ and $I_1$ associated with $\nu_e$ and $\bar\nu_e$. These labels are instrumental in discerning the presence or absence of $\nu$ELN  crossing.
It's worth highlighting that our current emphasis is primarily on the first two moments. These moments are of particular interest as they are the ones typically tracked directly in the simulation processes.

In order to train our ML algorithms, we partially employ 
two parametric neutrino angular distributions which have beed  widely used in the literature~\cite{Richers:2022dqa, Wu:2021uvt, Yi:2019hrp}, 
namely, 
 the maximum entropy distribution
defined as, 
\begin{equation}
f^{\rm{max-ent}}_\nu(\mu) = \exp[\eta + a\mu],
\end{equation}
and 
the Gaussian distribution,
\begin{equation}
f^{\rm{Gauss}}_\nu(\mu) = A\exp[-\frac{(1-\mu)^2}{a}],
\end{equation}
with,
\begin{equation}
 f_{\nu}(\mu) =  \int_0^\infty \int_0^{2\pi} \frac{E_\nu^2 \mathrm{d} E_\nu \mathrm{d} \phi_\nu}{(2\pi)^3} 
        f_{\nu}(\mathbf{p}).
\end{equation}
Here the parameters a, $\eta$,  and $A$ determines the overall neutrino number density
and  the shape of the neutrino distributions.

In addition, in order to improve the efficiency of our ML algorithms, we do feature engineering 
and instead of  $I_0$ and $I_1$ of  $\nu_e$ and $\bar\nu_e$ which are the provided information,
 we use,
\begin{equation}
\alpha = I^{\bar\nu_e}_0/I^{\nu_e}_0,\  F_{\nu_e} = I^{\nu_e}_1/I^{\nu_e}_0,\ \mathrm{and}\ F_{\bar\nu_e} = I^{\bar\nu_e}_1/I^{\bar\nu_e}_0,
\end{equation}
as the relevant features to be considered in the ML algorithms.
This is justified by bearing in mind that an overall normalisation factor does not affect the occurrence
of  $\nu$ELN crossings.

To facilitate the training and evaluation of ML algorithms, it is essential to partition the dataset into three distinct sets:
i) The training set: This subset is employed to train the ML algorithm, allowing it to learn patterns and relationships within the data,
ii) The development set: This set serves as a tool for fine-tuning the algorithm's hyperparameters, ensuring optimal performance, and
iii) The test set: This portion is dedicated to assessing the ML method's performance on previously unseen data, providing a reliable measure of its effectiveness. In practice, one should randomly distribute the data among these different datasets to avoid any sort of bias. In our analysis, we have randomly distributed the data  given the 0.8:0.1:0.1 fraction for  the training, development, and the test sets, respectively.

To comprehensively assess the effectiveness of our ML algorithm, we go beyond mere accuracy, which can be a somewhat simplistic measure. Instead, we consider  also  more detailed evaluations using precision and recall metrics, defined as,
\begin{equation}
\begin{split}
&\mathrm{accuracy} = \frac{T_p + T_n}{T_p + T_n + F_p + F_n} \\
&\mathrm{precision} = \frac{T_p}{T_p+F_p} \\
&\mathrm{recall} = \frac{T_p}{T_p+F_n} \\ 
&F_1 = 2\times \frac{\mathrm{precision} \times \mathrm{recall} }{\mathrm{precision} + \mathrm{recall}},
\end{split}
\end{equation}
with $T(F)_{p(n)}$ denoting True (False) positive (negative) classifications.
A discerning reader will notice that the precision/recall metric informs us about the reliability/detectability of  classifications, while $F_1$ is their harmonic mean. In this study, we opt for accuracy as the suitable metric because we aim for equal sensitivity to the presence or absence of $\nu$ELN crossings.

In this study we consider the following ML algorithms:
i) Logistic Regression (LR): a statistical classification algorithm that models the probability of a binary outcome. LR turns out to be one of the most promising ML algorithm to be used in detecting $\nu$ELN crossings~\cite{Abbar:2023kta},
ii) k-Nearest Neighbors (KNN): an intuitive  learning algorithm that classifies data points based on the majority class of their k-nearest neighbors in the feature space,
iii) Support Vector Machine (SVM):  a powerful algorithm that separates data into classes by finding the hyperplane that maximizes the margin between them in a high-dimensional space, and
iv) Decision Tree (DT): a tree-like model used for both classification and regression tasks, where the data is split into subsets based on feature conditions, ultimately leading to a decision or prediction.

There are two final aspect of the LR and SVM algorithms that requires a bit of clarification. 
While LR incorporates the nonlinear logistic function, it fundamentally operates as a linear classifier. Consequently, it cannot be directly applied to the detection of $\nu$ELN  crossings, which inherently represents a non-linear problem~\cite{Abbar:2023kta}.
To overcome this limitation, it becomes necessary to undertake a preprocessing step involving non-linear transformations and the creation of new features based on the original three features involved in the problem. The degree of polynomial transformation, being a hyper-parameter of this algorithm, plays a crucial role in this process~\footnote{The polynomial transformation method produces an augmented feature matrix by considering all possible polynomial combinations of the original features up to a specified degree. To illustrate, if an input sample has two dimensions denoted as $(a, b)$, the resulting degree-2 polynomial features would include $(a, b, a^2, ab, b^2)$ (assuming no bias term is taken into account).}.

Regarding the SVM algorithm, we should mention that we employ the radial basis function (RBF) kernel, defined as,
$\mathcal{K}(x,x') = \exp(-\gamma ||x-x'||^2)$.
 Here $\gamma$ is a hyper-parameter which is set to be $\gamma = 100$~\cite{Abbar:2023kta}.
%The RBF kernel is based on finding similarities between the test data point and the ones in the training set, where  $\gamma$ 
% determines the  distance up to which the similarities can efficiently make an impact.
 
In the next part, we present our results. In addition, to promote transparency and collaboration, we have made our ML methodologies available on \href{https://github.com/sajadabbar/realistic-data-XLN.git}{GitHub}.

  \subsection{ML-based detection of $\nu$ELN crossings using the SN data}
  
   \begin{table}[b]
\centering
\caption{A summary of the metric scores of the previously-trained ML algorithms (using artificial data) tested on the realistic dataset. This is to be compared with 
TABLE I. in Ref.~\cite{Abbar:2023kta}. Alongside each algorithm, one can find its corresponding accuracy score.}
\begin{tabular}[t]{|lcc|c|}
\hline
& \textcolor{black}{ \textbf{{LR} (n = 9)} (68\%)}    \\
\hline
& precision & recall & $F_1$-score \\
\hline
no  crossing & 72\% & 82\% & 77\% \\
 crossing&59\%&43\% & 50\% \\
\hline
&\textcolor{black}{  \textbf{{KNN (n=3)}} (77\%)  } \\
\hline
&precision&recall & $F_1$-score \\
\hline
no  crossing&77\%&89\%&83\%\\
 crossing&75\%&55\%&63\%\\
\hline
&\textcolor{black}{   \textbf{{SVM}} (87\%)}\\
\hline
&precision&recall & $F_1$-score\\
\hline
no  crossing&98\%&81\%&89\%\\
 crossing&75\%&98\%&85\%\\
\hline
&\textcolor{black}{   \textbf{{Decision tree}} (71\%) }\\
\hline
&precision&recall & $F_1$-score \\
\hline
no  crossing&74\%&84\%&79\%\\
 crossing&63\%&48\%&55\%\\
\hline
\end{tabular}
\label{tab:tab1}
\end{table}%

\begin{figure} [tb]%  figure placement: here, top, bottom, or page
\centering
\begin{center}
\includegraphics*[width=.45\textwidth, trim= 0 0 30 20, clip]{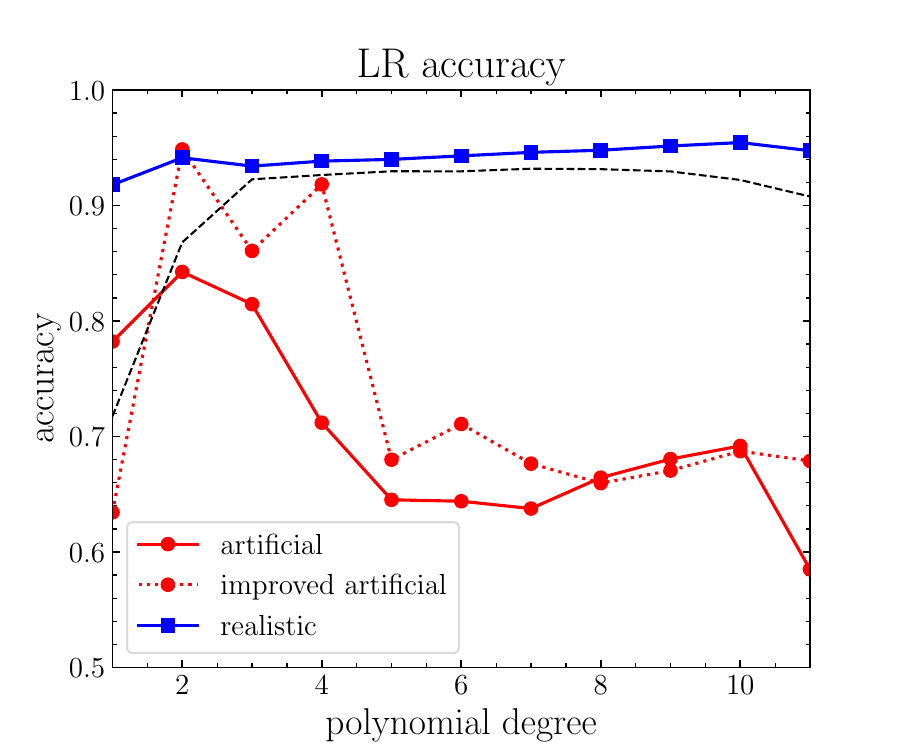}
\end{center}
\caption{
The accuracy of the LR algorithms  evaluated on the realistic dataset for models trained on various training sets, with a focus on the impact of polynomial degree of the nonlinear transformations. It is noteworthy that LR models trained on the improved artificial training sets can achieve comparable accuracies to those trained on realistic data at lower polynomial degrees, implying simpler model structures.
The black dashed line represents the performance of the old LR model trained on artificial data when tested on its dedicated test set.}
\label{fig:LR_1}
\end{figure}

 We initially assess the performance of the pre-trained ML models of Ref.~\cite{Abbar:2023kta} (using artificial data) on the realistic dataset obtained from the simulation. The metric scores for the performance of these pre-trained ML algorithms  are presented in TABLE.~\ref{tab:tab1}. It becomes evident that the performance of ML models, which were initially trained on artificial data, degrades when applied to realistic data. Notably, the accuracies of  LR and SVM are of paramount concern. Specifically, the LR model (here was considered with a polynomial degree of 9) exhibits a significant deviation from its accuracy on artificial data (as indicated in TABLE I of Ref.~\cite{Abbar:2023kta}). This indicates poor generalization of the old LR algorithm with a polynomial degree $n=9$, a known issue referred to as high variance in the context of ML. 
 
Another enlightening aspect centers on the SVM model's relatively good performance, consistently achieving high accuracy even when confronted with previously unseen data. This serves as yet another compelling example of the effectiveness of maximizing margins to separate classes, underscoring its conceptual soundness and its ability to enhance the model's overall generalizability in ML.

However and  as depicted in Fig.~\ref{fig:LR_1}, the former LR model (red-curve) trained on synthetic data, attains its peak performance around $n\simeq2$ achieving an accuracy of approximately 85\%. In simpler terms, the more straightforward models outperform their counterparts when it comes to $\nu$ELN detection on previously unseen data. This serves as a prime illustration of the bias-variance tradeoff within the realm of ML. As we see later on, it seems to be  a general observation that when considering LR, 
opting for nonlinear transformations with lower polynomial degrees tends to yield better results, at least as long as  a 
variation between the training and test sets is expected.
 
 % Though the accuracy of  $\sim85\%$ could be still considered impressive given the fact that
% realistic angular distributions might be very different from the artificial ones, one can still improve the performance 
% of the models developed using artificial dataset. Indeed, the old ML models were trained on the data with 
 % $\alpha$'s  in the range $\alpha = (0.03 - 2.5)$.  
% Then for each $\alpha$, a number of random $F_{\nu_e}$ and $F_{\bar\nu_e} $ in range~(0,1) were considered.
 %Though this was ok for first step, this does not reflect what one expects in realistic conditions. Instead, in realistic
% SN and NSM merger simulations one expects  $F_{\nu_e} \lesssim F_{\bar\nu_e} $. This means only a portion of the 
% parameter space is covered in the plane of  $F_{\nu_e} \lesssim F_{\bar\nu_e} $. 

 While achieving an accuracy of $\sim85\%$ is already acceptable, it's important to note that the performance of models developed using artificial datasets can still be enhanced. The initial ML models were trained on data with $\alpha $ values ranging from 0.03 to 2.5, alongside random selections of $F_{\nu_e}$ and $F_{\bar\nu_e} $  in the (0, 1) range. However, this approach, while suitable for an initial step, doesn't align with realistic conditions. In realistic simulations of CCSNe, it's anticipated that $F_{\nu_e} \lesssim F_{\bar\nu_e} $ (note that this could be a bit distorted in the case of NSM where the radiation fields can be qualitatively different).

To address this issue, we enhance our ML model by training it with artificial data while considering $F_{\nu_e}$ within the range of $(0.6 F_{\bar\nu_e}, F_{\bar\nu_e})$. The performance of such a ML model on realistic data is illustrated in Fig.~\ref{fig:LR_1} (red-dotted curve), revealing two significant insights. Firstly, high variance is observed at large polynomial degrees, indicating poor generalizability of the ML performance at those polynomial degrees. Secondly, the ML model, trained on this improved artificial dataset, achieves notably higher accuracy compared to the previous version. This reaffirms the significance of a well-representative training dataset for accurate testing.

Hence, variations in the distributions between training and test datasets can  impact accuracy. It's essential to emphasize that this scenario differs entirely from encountering unexplored regions of the parameter space in the test set.
To better understand this, let's consider an ML classifier trained on a dataset comprising 500 cat images, 490 horse images, and only 10 dog images, achieving an overall accuracy of 99\%. However, the accuracy drops to 30\% when classifying dog images. This discrepancy arises because most of the training focus was directed towards enhancing the classification of more abundant cat and horse images.
Now, envision a test set consisting of 500 cat images and 500 dog images. Despite some dog images being present in the training set, a subpar performance on the test set is anticipated due to differing distributions compared to training set.

While this approach notably improves the ML model's performance in CCSN environments, the constrained parameter space within the training set could also partially restricts the applicability of the ML model to that specific parameter range.

\begin{figure} [tb]%  figure placement: here, top, bottom, or page
\centering
\begin{center}
\includegraphics*[width=.45\textwidth, trim= 0 0 30 20, clip]{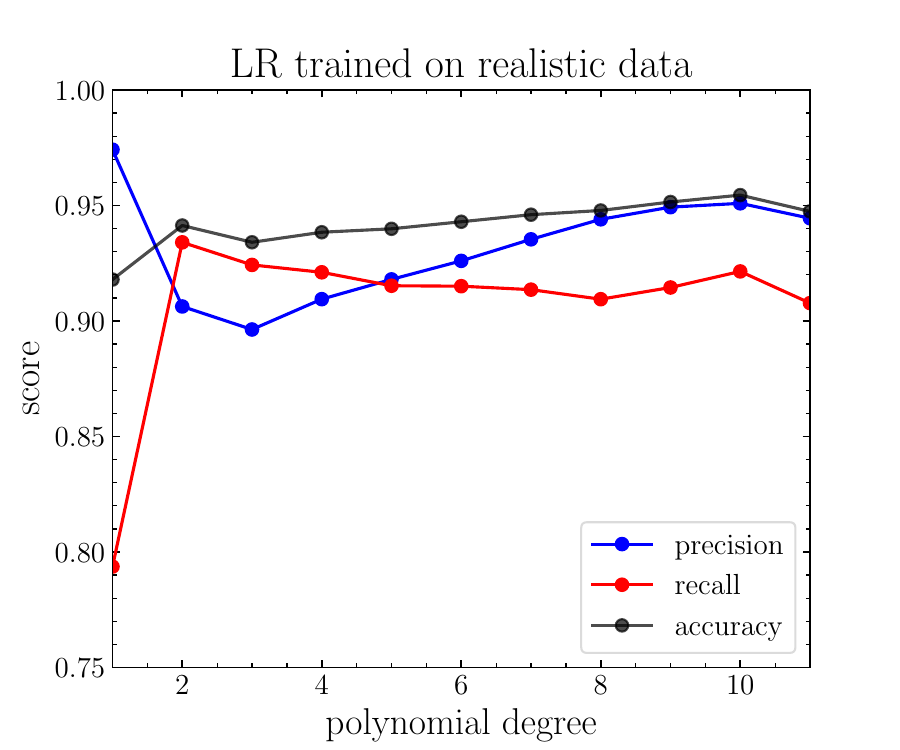}
\end{center}
\caption{
The metric scores of the LR algorithm trained on a combination of the artificial and realistic datasets, as a function of the 
polynomial degree of the nonlinear transformations. 
It's worth noting that the precision and recall scores typically exhibit opposing trends, a phenomenon commonly referred to as the precision-recall tradeoff within the field of ML.}
\label{fig:LR_real}
\end{figure}

In the final step, we enhance our ML model by incorporating real-world data into the training set. It's important to highlight that we don't exclusively rely on real data for training. This approach allows us to maintain variability in the angular distributions of neutrinos. The performance of this ML model is depicted by the blue curve in Fig.~\ref{fig:LR_1}. Notably, it exhibits exceptionally high accuracy across all polynomial degrees. However, drawing from our previous experiences, we favor selecting $n=2$ due to its strong potential for effective generalization to unseen data. Beyond its generalization capabilities, opting for $n=2$ also brings the advantage of reduced computational intensity when implementing the LR model in CCSN and NSM simulations on the fly.
In addition, In Fig.~\ref{fig:LR_real}, we present the comprehensive set of metric scores obtained from our improved LR model. It's worth highlighting that all these scores attain satisfactory values when the polynomial degree is set to 2.

   \begin{table}[tb]
\centering
\caption{A summary of the metric scores of ML algorithms trained on the combination of the realistic and artificial datasets, and then tested with the realistic data. Alongside each algorithm, one can find its corresponding accuracy score.}
\begin{tabular}[t]{|lcc|c|}
\hline
& \textcolor{black}{ \textbf{{LR} (n = 2)} (94\%)}    \\
\hline
& precision & recall & $F_1$-score \\
\hline
no  crossing & 96\% & 95\% & 95\% \\
 crossing&91\%&93\% & 92\% \\
\hline
&\textcolor{black}{  \textbf{{KNN (n=3)}} (98\%)  } \\
\hline
&precision&recall & $F_1$-score \\
\hline
no  crossing&98\%&99\%&99\%\\
 crossing&98\%&97\%&98\%\\
\hline
&\textcolor{black}{   \textbf{{SVM}} (97\%)}\\
\hline
&precision&recall & $F_1$-score\\
\hline
no  crossing&98\%&98\%&98\%\\
 crossing&96\%&97\%&97\%\\
\hline
&\textcolor{black}{   \textbf{{Decision tree}} (99\%) }\\
\hline
&precision&recall & $F_1$-score \\
\hline
no  crossing&99\%&99\%&99\%\\
 crossing&98\%&98\%&98\%\\
\hline
\end{tabular}
\label{tab:tab2}
\end{table}%

In Table~\ref{tab:tab2}, we present the performance results of our enhanced ML models, which were trained using a combination of both realistic and artificial datasets. The outcomes demonstrate that our ML models achieve very high performance metrics. Notably, these scores surpass those obtained by the ML model trained exclusively on artificial data, as  shown in Table I of Ref.~\cite{Abbar:2023kta}. This improvement can be attributed to the fact that
the artificial data includes noisy labels, as discussed in Ref.~\cite{Abbar:2023kta}, which was identified as a primary source of inaccuracies
once the ML model is tested on artificial data.

In all our calculations, we've primarily employed random data distribution between training and test datasets. While this is crucial to ensure having a robust ML model, we conducted also additional calculations where the training and testing processes were executed on datasets belonging to distinct time snapshots. We observed that these variations yielded similar results, confirming the robustness of our methods.

\subsection{Performance of ML in the neutrino decoupling region}

In the previous section, we assessed the overall performance of ML in the  SN environment. It's important to note that FFCs are expected to be most impactful when they occur closer to the surface of the PNS, well within the SN post-shock zone. This can be attributed to two primary reasons. Firstly, when FFCs happen in deeper SN regions, the $\nu$ELN crossings can potentially be wider/deeper, which may result in more pronounced flavor conversions. Additionally, any flavor conversion occurring above the SN shock is not expected to have a noticeable impact on the CCSN dynamics~\cite{Ehring:2023abs}, though can still impact the neutrino signal.

Considering this, in addition to evaluating our ML methods' overall performance, we specifically examined their performance in SN regions located well below the shock. To do this, we assessed the performance of our ML algorithms in SN zones where the radial distance  is $\lesssim 100$~km. As illustrated in Table~\ref{tab:tab_dec}, our ML models effectively do a good job in capturing $\nu$ELN crossings. The only exception is the recall score of the  LR method, which is relatively low at around 50\%, despite  its  good overall performance.

\begin{figure} [tb]%  figure placement: here, top, bottom, or page
\centering
\begin{center}
\includegraphics*[width=.45\textwidth, trim= 0 30 30 50, clip]{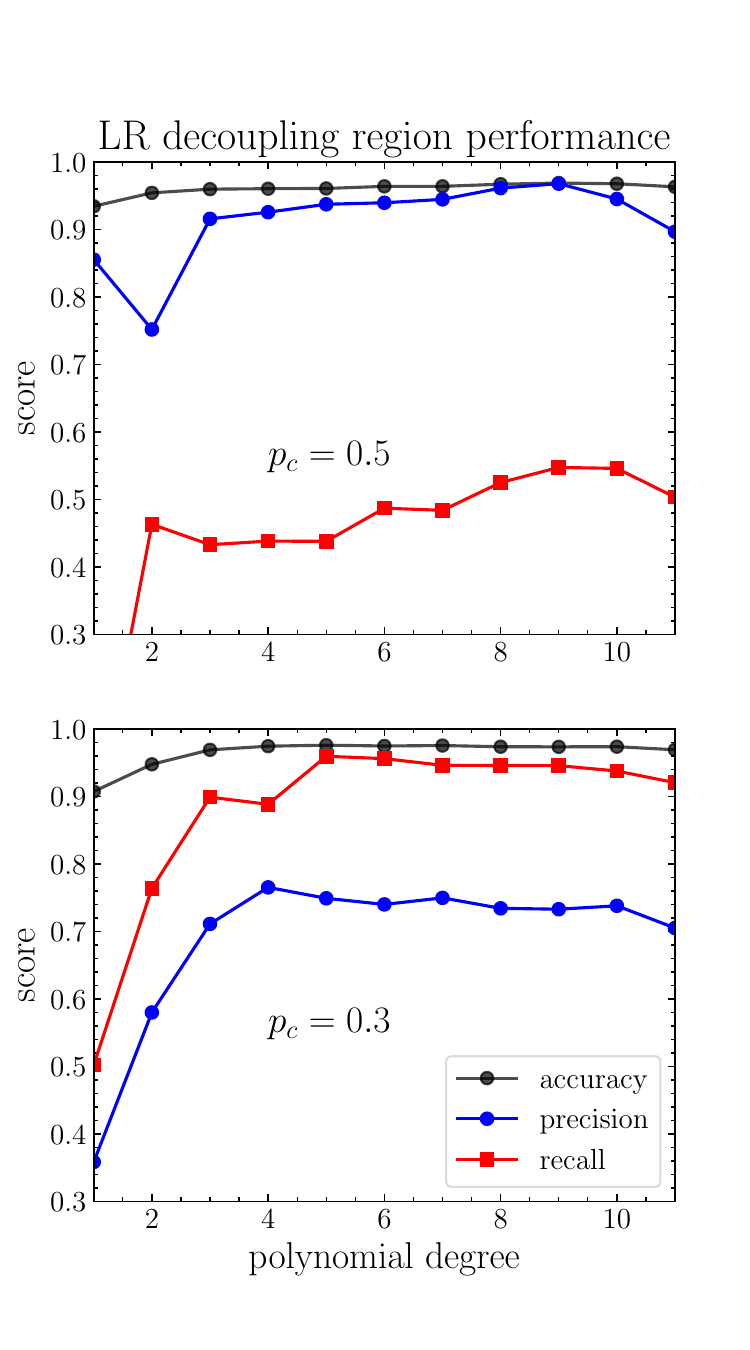}
\end{center}
\caption{
The performance of the LR algorithm in the neutrino decoupling region, as a function of the 
polynomial degree of the nonlinear transformations. The LR is trained on a combination of the artificial and realistic datasets.
As mentioned previously, it's worth noting that the precision and recall scores typically exhibit opposing trends due to precision-recall tradeoff. As one can see in the lower panel, one can enhance the recall score significantly by decreasing 
the threshold probability of LR, $p_c$, though with the price
of a reduction in the precision score.}
\label{fig:LR_decoup}
\end{figure}

Fixing this issue involves lowering the threshold probability (denoted as $p_c$) for the LR algorithm, as discussed in Ref.~\cite{Abbar:2023kta}. 
As depicted in Fig.~\ref{fig:LR_decoup}, significant improvement of the recall score can be achieved by reducing the LR threshold probability (here to $p_c=0.3$). However, this comes at the cost of a reduction in the precision score, showcasing the traditional precision-recall tradeoff. While other ML algorithms exhibit very high scores, rendering them ideal for detecting the post-processing 
detection of $\nu$ELN crossings,
  the LR algorithm remains valuable due to its ease of implementation in CCSN and NSM simulations. Therefore, we believe it is still worth considering the LR method and adjusting its threshold probability, despite the inherent tradeoff between precision and recall.

  \begin{table}[tb!]
\centering
\caption{A summary of the metric scores of  ML algorithms trained on the combination of the realistic and artificial datasets,
in the SN post-shock region, namely at radii~$\lesssim 100$~km. %This is to be compared with TABLE I. in Ref.~\cite{}. 
Alongside each algorithm, one can find its corresponding accuracy score.}
\begin{tabular}[t]{|lcc|c|}
\hline
& \textcolor{black}{ \textbf{{LR} (n = 2)} (96\%)}    \\
\hline
& precision & recall & $F_1$-score \\
\hline
no  crossing & 96\% & 100\% & 98\% \\
 crossing&94\%&49\% & 64\% \\
\hline
&\textcolor{black}{  \textbf{{KNN (n=3)}} (100\%)  } \\
\hline
&precision&recall & $F_1$-score \\
\hline
no  crossing&100\%&100\%&100\%\\
 crossing&100\%&99\%&99\%\\
\hline
&\textcolor{black}{   \textbf{{SVM}} (99\%)}\\
\hline
&precision&recall & $F_1$-score\\
\hline
no  crossing&99\%&99\%&99\%\\
 crossing&92\%&90\%&91\%\\
\hline
&\textcolor{black}{   \textbf{{Decision tree}} (100\%) }\\
\hline
&precision&recall & $F_1$-score \\
\hline
no  crossing&100\%&100\%&100\%\\
 crossing&99\%&99\%&99\%\\
\hline
\end{tabular}
\label{tab:tab_dec}
\end{table}%

  \subsection{Detection of $\nu$ELN-XLN crossings}

So far, our discussion has focused exclusively on detecting $\nu$ELN crossings. However, in a broader context, it's important to acknowledge that the angular distributions of $\nu_x$ and $\bar\nu_x$ can be different in CCSN and NSM environments. This difference  becomes particularly pronounced when we account for the potential creation of muons at the core of these extreme astrophysical objects~\cite{Bollig:2017lki, Fischer:2020vie, Guo:2020tgx}.
Consequently, to accurately identify the occurrence of FFCs under the most realistic conditions, we must shift our attention towards detecting $\nu$ELN-XLN crossings rather than confining ourselves to the $\nu$ELN ones.

The distinction between the detection of $\nu$ELN-XLN crossings and $\nu$ELN crossings presents several key differences. One of the most significant distinctions lies in the increased complexity of required information. In the context of our ML methods, this translates to an expansion in the number of essential features. Specifically, we now necessitate seven features instead of the previous three, namely $\alpha_{\nu_e}$,  $\alpha_{\nu_x}$, $\alpha_{\bar\nu_x}$, 
$F_{\nu_e}$, $F_{\bar\nu_e}$, $F_{\nu_x}$, and $F_{\bar\nu_x}$, with $\alpha_{\nu_\beta}=n_{\nu_\beta}/n_{\nu_e}$.

The increase of the number of features and the greater demand for information significantly contribute to an elevated classification error in this context. Notably, when $\nu_x$ and $\bar\nu_x$ exhibit disparities, the $\nu$ELN-XLN profile can exhibit more intricate characteristics, e.g., it is conceivable that even multiple crossings occur.

   \begin{table}[tb]
\centering
\caption{A summary of the metric scores of the  ML algorithms for $\nu$ELN-XLN crossing detection. %This is to be compared with TABLE I. in Ref.~\cite{}. 
Alongside each algorithm, one can find its corresponding accuracy score.}
\begin{tabular}[t]{|lcc|c|}
\hline
& \textcolor{black}{ \textbf{{LR} (n = 2)} (88\%)}    \\
\hline
& precision & recall & $F_1$-score \\
\hline
no  crossing & 87\% & 89\% & 88\% \\
 crossing&88\%&86\% & 87\% \\
\hline
&\textcolor{black}{  \textbf{{KNN (n=3)}} (88\%)  } \\
\hline
&precision&recall & $F_1$-score \\
\hline
no  crossing&89\%&88\%&89\%\\
 crossing&88\%&89\%&88\%\\
\hline
&\textcolor{black}{   \textbf{{SVM}} (88\%)}\\
\hline
&precision&recall & $F_1$-score\\
\hline
no  crossing&92\%&84\%&88\%\\
 crossing&85\%&93\%&89\%\\
\hline
&\textcolor{black}{   \textbf{{Decision tree}} (87\%) }\\
\hline
&precision&recall & $F_1$-score \\
\hline
no  crossing&87\%&87\%&87\%\\
 crossing&87\%&87\%&87\%\\
\hline
\end{tabular}
\label{tab:tab3}
\end{table}%

 In order to train our ML models, we use artificial distributions for  neutrinos, given the fact that labeled
 data regarding the existence of $\nu$ELN-XLN crossings are not available. 
 In order to prepare our data,  we consider $\alpha_{\nu_e}$ $\alpha_{\nu_x (\bar\nu_x)}$ to be in the range
 of (0., 2.5) and (0., 3.), respectively. Also we assume an allowed  maximum 40\% difference between
$ \nu_x $ and $\bar\nu_x$ quantities. This is consistent with the observation that the difference between 
$ \nu_x $and $\bar\nu_x$ should be subdominant in realistic simulations~\cite{Bollig:2017lki}.
 In addition, we keep in mind what we learned previously that the data should be enough representative of the
 realistic data. Thus, we also respect the hierarchy $F_{\nu_e} \lesssim F_{\bar\nu_e}  \lesssim F_{\nu_x (\bar\nu_x)}$.

The performance of our ML models in detecting $\nu$ELN-XLN crossings is presented in Table~\ref{tab:tab3}. Notably, the overall performance lags behind that of $\nu$ELN crossing detection. This disparity can be attributed to the presence of intricate patterns governing the crossings and an increase in label noise.

\section{DISCUSSION AND OUTLOOK}
Recent advancements have showcased the remarkable capabilities of ML in identifying the $\nu$ELN crossings in the CCSN and NSM simulations~\cite{Abbar:2023kta}. In this study, we have propelled prior research in two pivotal and distinctive directions. Firstly, we have subjected ML models to the rigorous test of real-world data acquired from CCSN simulations, where the intricate problem of neutrino transport is addressed through the comprehensive Boltzmann equation. Secondly, we have expanded our ML techniques to encompass the detection of $\nu$ELN-XLN crossings, accommodating situations where there may exist distinctions between $\nu_x$ and $\bar\nu_x$.

Using realistic CCSN data, we have demonstrated that the simpler models consistently outperform their more complex counterparts in the context of $\nu$ELN detection when applied to previously unseen data. This provides a clear and compelling example of the bias-variance tradeoff within the domain of ML.
Specifically, it was observed that the LR model performs most effectively when a polynomial transformation of degree $n=2$ is applied, as opposed to the previously suggested degree of $n=9$ which was based on artificial data. This underscores the importance of considering the complexity of models and their suitability for real-world data, highlighting that sometimes, a simpler approach can yield superior results.

We demonstrate a significant enhancement in model performance when utilizing artificial datasets by aligning the parameter space of the synthetic data with the realism expected in CCSN and NSM simulations. Specifically, we adhere to the hierarchy $F_{\nu_e} \lesssim F_{\bar\nu_e} \lesssim F_{\nu_x (\bar\nu_x)}$, which mirrors the conditions anticipated in these astrophysical events.
This deliberate consideration in the preparation of artificial data results in model performance that rivals that of ML models trained on realistic data, at least within a certain parameter range.

We have further fortified our ML models by integrating real-world data into the training set, and this enhancement has yielded remarkable generic accuracy in out ML models. Based on our  observations, we are inclined to assert that the LR model with a polynomial degree of $n=2$ stands out as the optimal choice for detecting FFCs in CCSN and NSM simulations. This choice not only exhibits effective generalization capabilities for unseen data but also offers the distinct advantage of reducing computational overhead when deploying the LR model in real-time CCSN and NSM simulations.

We have also developed ML models to identify neutrino flavor crossings in the $\nu$ELN-XLN distributions. This is particularly relevant because the angular distributions of $\nu_x$ and $\bar\nu_x$ can exhibit variations in CCSN and NSM environments.
Unlike the simpler task of detecting $\nu$ELN crossings, detecting $\nu$ELN-XLN crossings introduces a higher level of complexity due to the need for more  information. In the context of our ML methods, this complexity manifests as an increase in the number of essential features. This augmented feature set  substantially contribute to an elevated classification error in this specific context.

 In summary, our study significantly expands upon prior research, allowing for more confident utilization of ML methods in detecting FFCs. However, there remain crucial avenues for exploration. Specifically, our current analysis focuses on crossings occurring in the zenith angle ($\mu$), when the angular distribution is integrated over $\phi_\nu$. Nevertheless, as demonstrated in previous references, a substantial fraction of $\nu$ELN crossings may be exclusively in $\phi_\nu$, exhibiting non-axisymmetric behavior~\cite{Nagakura:2019sig}. Thus, it is imperative to develop ML techniques capable of capturing these non-axisymmetric crossings. Closely related to this issue is the exploration of FFCs in rotating CCSN models~\cite{Harada:2021ata}, as the prominence of non-axisymmetric features in such models could influence the performance of ML algorithms. 
 Given the proven and remarkable versatility and effectiveness of ML in this context, implementing these measures will further improve the detection of FFCs in CCSN and NSM simulations.

%%%%%%%%%%%%%%%%%%%%%%%%%%%%%%%%%%%%%%%%%%%%%%%%%%%%%%%%%%%%%%%%%%%%%%%%%%%%%
% Place all of the references you used to write this paper in a file
% with the same name as following the \bibliography command
%%%%%%%%%%%%%%%%%%%%%%%%%%%%%%%%%%%%%%%%%%%%%%%%%%%%%%%%%%%%%%%%%%%%%%%%%%%%%

\section*{Acknowledgments}
We would like to thank Georg Raffelt for useful discussions. 
We would also like to express our sincere gratitude to the Institute of Physics of Academia Sinica for their warm hospitality and support during the \emph{Focus Workshop on Collective Oscillations and Chiral Transport of Neutrinos}, where the inception of this project took place. Their gracious hosting and collaborative environment played a pivotal role in shaping the foundation of our work.
S.A. was supported by the German Research Foundation (DFG) through
the Collaborative Research Centre  ``Neutrinos and Dark Matter in Astro-
and Particle Physics (NDM),'' Grant SFB-1258, and under Germany’s
Excellence Strategy through the Cluster of Excellence ORIGINS
EXC-2094-390783311.
H. N. was supported by Grant-inAid for Scientific Research (23K03468) and also by the NINS International Research Exchange Support Program. The numerical simulations for CCSNe were carried out by using "K", ”Fugaku”, and the highperformance computing resources of ”Flow” at Nagoya University ICTS through the HPCI System Research Project (Project ID: 220173, 220047, 220223, 230033, 230204, 230270).
We would also like to acknowledge the use of the following softwares: \textsc{Scikit-learn}~\cite{pedregosa2011scikit}, \textsc{Matplotlib}~\cite{Matplotlib}, \textsc{Numpy}~\cite{Numpy}, \textsc{SciPy}~\cite{SciPy}, and \textsc{IPython}~\cite{IPython}.

\bibliographystyle{elsarticle-num}%apsrev4-1}
\bibliography{Biblio}
% \bibliography{ref}

%%%%%%%%%%%%%%%%%%%%%%%%%%%%%%%%%%%%%%%%%%%%%%%%%%%%%%%%%%%%%%%%%%%%%%%%%%%%%
\clearpage
%\appendix

\end{document}